\documentclass[twocolumn,showpacs,preprintnumbers,amsmath,amssymb]{revtex4}
\usepackage{graphicx}
\usepackage{dcolumn}
\usepackage{bm}

\begin{document}

\title{Twisting and Bending Stress in DNA Minicircles }

\author{ Marco Zoli }
\affiliation{
School of Science and Technology - CNISM \\  Universit\`{a} di Camerino, I-62032 Camerino, Italy \\ marco.zoli@unicam.it}

\date{\today}

\begin{abstract}
The interplay between bending of the molecule axis and appearance of disruptions in circular DNA molecules, with $\sim 100$ base pairs, is addressed.
Three minicircles with different radii and almost equal content of AT and GC pairs are investigated.
The DNA sequences are modeled by a mesoscopic Hamiltonian which describes the essential interactions in the helix at the level of the base pair and incorporates twisting and bending degrees of freedom. Helix unwinding and bubble formation patterns are consistently computed by a path integral method that sums over a large number of molecule configurations compatible with the model potential. The path ensembles are determined, as a function of temperature, by minimizing the free energy of the system. Fluctuational openings appear along the helix to release the stress due to the bending of the molecule backbone.  In agreement with the experimental findings, base pair disruptions are found with larger probability in the smallest minicircle of \textit{66-bps} whose bending angle is $\sim 6^{ o} $. For this minicircle,  a sizeable untwisting is obtained with the helical repeat showing a  step-like increase at $\tau =\,315K$. The method can be generalized to determine the bubble probability profiles of open ends linear sequences.
\end{abstract}


\maketitle

\section*{I. Introduction}

Fundamental methods of molecular biology, such as probe target hybridization and polymerase chain reaction, depend on the knowledge of thermal stability of short DNA duplexes with specific sequences \cite{benight,genzer}. Accordingly, large scale implementation of these techniques requires reliable algorithms, generally based on nearest-neighbors statistical thermodynamics, to predict the oligomers melting profiles in various ambient conditions  \cite{santa,blake,owc,rapti10}. Starting from the peculiar Watson-Crick hybridization rule, uncountable DNA molecules and architectures have been designed over the last years for application in functional nanodevices and drug deliveries
\cite{seeman,rose,zhao,cha,albu}.

Curved molecules can be identified by their slow electrophoretic mobility \cite{crothers,shore}. Being common to the control regions of transcription, the bending of the double helix is essential to gene regulation and to the compaction of genomic DNA into chromatin \cite{ohyama}. Short sequences offer remarkable theoretical challenges and permit to probe DNA at the scale of the genetic code \cite{choi}.
One may expect that fragments of about $100$ base pairs (\emph{bps}), being shorter than a persistence length ($\sim 500${\AA}), have a large stiffness which prevents them from closing into a circle. However, Cloutier and Widom  \cite{cloutier} measured a much larger cyclization probability \cite{stock} than that predicted by the traditional worm-like-chain model \cite{shimada} thus suggesting that bending may occur spontaneously also in short fragments, even in the absence of proteins, due to an intrinsic DNA flexibility.
Besides, the loop formation probability appeared to be very sensitive to the molecular length and sequence. To interpret these findings, Yan and Marko proposed \cite{yan04} that the opening of localized bubbles may be the mechanism which energetically favors the cyclization of short molecules.   In fact, Crick and Klug \cite{crick} had first put forward the idea that the unstacking of two adjacent \emph{bps} along the helix axis, a kink, may facilitate strong bends of the molecule albeit maintaining the hydrogen bond base pairing \cite{maddocks}. While kinks can reduce the bending energy and enhance the cyclization probability \cite{volo05}, they have a lower entropy than that associated to the \emph{bps} breaking \cite{zocchi13} which, accordingly, confers a higher flexibility to the molecule and also changes the helical repeat.
More recently, single-strand-specific endonucleases experiments have shown \cite{volo08} that small circles of different sizes react differently to the bending stress: the latter may cause local disruption of the double helix  (either in the form of kinks or base pair breaking) in $\sim 66$-\emph{bps} circles whereas disruptions are not detected in larger circles, $\sim  86$-\emph{bps} and $\sim 106$-\emph{bps}. The circle sizes have been chosen in order to have a number of helix turns slightly larger than an integer number in order to reduce the effects of the torsional stress and emphasize the role of the bending on the helix conformation.
These experimental findings have been justified by a Monte-Carlo simulation of a discrete worm-like chain \cite{volo09} (with adjustable bending potential parameters) in which the disruption probability has been taken independent of the sequence site. After years of experimental and theoretical research \cite{vafa,volo13}, it remains however unsettled whether there exists a critical radius of curvature which causes a base pair opening in short circles.

These issues are examined in this paper from a different viewpoint. Here we apply path integral techniques \cite{i11} to a mesoscopic Hamiltonian model which accounts for the sequence specificities at the level of the base pair and incorporates the rotational degrees of freedom, i.e. both the base pair twisting around the molecule axis and the bending of the axis itself. In particular, we focus on the interplay between bending and base pair breaking by monitoring the appearance of isolated bubbles, as a function of the molecule length, in three DNA minicircles analyzed in ref.\cite{volo08}.  Unlike previous studies, we don't assume that the helical repeat ($h$) gets a constant value independent of the circle size, accepting instead that significant variations in $h$ may be found among the various systems.
In general, base pair disruptions changing locally the helix conformation may be averaged out in long chains thus leaving scarce traces on $h$ which is an average parameter of the molecule \cite{duguet}. This is however not the case in short sequences where such local deformations have an impact on the value of $h$.
Accordingly, in our analysis, the room temperature equilibrium conformation is determined, for each minicircle, as the one which minimizes the free energy after computing the latter throughout a broad range of $h$ values. Furthermore, looking at the thermal effects on the helix conformations, we admit that $h$ may depend on temperature ($\tau$)  as experimentally envisaged since long \cite{depew,benham93b} and re-determine it for each investigated sequence, at any $\tau$, on the base of the minimum energy criterion. Implementation of this program is preliminary to an accurate computation of the bubble probability profiles as local denaturation and bubble formation are, \textit{in vivo}, regulated by the degree of helical stress \cite{benham99,metz12}. Thus, torsional and bending stresses appear crucial
to select those sites along the sequence which are more sensitive to disruption phenomena.

\section*{II. Sequences of Circular DNA}

We consider three minicircles of different sizes and almost equal content of GC-\textit{bps} and AT-\textit{bps}, whose preparation methods and sequences are given in ref.\cite{volo08} with Supplementary Data.

\begin{eqnarray}
& &\textbf{106 bps} \, \nonumber
\\
& &ATCTTTGCGGCAGTTAATCGAACAAGACCC \, \nonumber
\\
& &GTGCAATGCTATCGACATCAAGGCCTATCG \, \nonumber
\\
& &CTATTACGGGGTTGGGAGTCAATGGGTTCA \, \nonumber
\\
& &GGATGCAGGTGAGGAT \, \nonumber
\\
\, \nonumber
\\
& &\textbf{86 bps} \, \nonumber
\\
& &ATCTTCATCGAACAAGACCCGTGCAATGCT \, \nonumber
\\
& &ATCGACATCAAGGCCTATCGCTTGGGAGTC \, \nonumber
\\
& &AATGGGTTCAGGATGCAGGTGAGGAT \, \nonumber
\\
\, \nonumber
\\
& &\textbf{66 bps} \, \nonumber
\\
& &ATCTTATCGCGTGCAATGCTATCGACATCA \, \nonumber
\\
& &AGGCCTATCGCTGGGAGTCAATGAGCAGGT \, \nonumber
\\
& &GAGGAT \, \nonumber
\\
\label{eq:10}
\end{eqnarray}

In view of the shortness of these circles, it can be reasonably assumed that the helix axis lies in a plane \cite{bates} hence, the writhe number which involves
crossings of the helix axis over itself vanishes, $Wr=\,0$. Then, the DNA supercoiling, measured by the linking number $Lk$, is due only to the twist $Tw$ of the individual strands around the helical axis, $Lk=\,Tw$ \cite{shore83,marko95}.  The helical repeat is $h=\, N / Tw$, $N$ being the number of \textit{bps} forming the circle.

\section*{III. Mesoscopic model}

While fully atomistic approaches are computationally intractable even in short sequences due to the huge number of degrees of freedom, mesoscopic models have the advantage to capture the main interactions at play in the helix maintaining a description in terms of the base pairs.
We use a Dauxois-Peyrard-Bishop Hamiltonian  \cite{pey2} generalizing it to a circle of radius $R$.  Say $\textbf{r}_i$, the inter-strand base pair displacement with respect to the ground state. The latter is found when all \textit{bps} centers of mass are uniformly arranged on the circumference which represents the molecule backbone.
Then, we define the modulus $\eta _i$ of the base pair fluctuational vector:

\begin{eqnarray}
& &\bigl({\eta }_i \bigr)_{x} =\, |\textbf{r}_i| \cos\phi_i \cos\theta_i \, \nonumber
\\
& &\bigl({\eta }_i\bigr)_{y} =\,(R + |\textbf{r}_i|\sin\theta_i) \cos\phi_i
\, \nonumber
\\
& &\bigl({\eta }_i\bigr)_{z} =\,(R + |\textbf{r}_i|) \sin\phi_i \,.
\,
\label{eq:1}
\end{eqnarray}

The ground state is recovered once all \textit{bps}-fluctuations vanish hence, $\eta_i =\,R ,\,\, \forall i$.
The polar angle, $\theta_i =\, (i - 1) 2\pi / h + \theta_S$, measures the $i-$ \emph{bp} twisting around the molecule backbone.
$\theta_S$ is the twist of the first \emph{bp} from the left in each of the sequences in Eq.~(\ref{eq:10}).
The azimuthal angle, $\phi_i =\, (i-1){{2 \pi} / N}$,  measures the bending between adjacent \emph{bps} along the stack.
The fluctuational orbits defined by $i=\,1$ and $i=\,N+1$ overlap consistently with the closure condition holding for the DNA ring.
A detailed description of our circular model is given in ref. \cite{i13}.  The mesoscopic Hamiltonian consists of the following terms:

\textit{a)} a Morse potential

\begin{eqnarray}
V_M[\eta_i]=\,D_i \bigl[\exp(-b_i (\eta_i - R)) - 1 \bigr]^2 \, ,
\label{eq:6a}
\end{eqnarray}

simulating the hydrogen bond between the $i-th$ base pair mates on the complementary strands. $D_i$ and $b_i$ are the site dependent pair breaking energy and inverse length, respectively.

\textit{b)} a solvent potential

\begin{eqnarray}
V_{sol}[\eta_i]=\, - D_i f_s \bigl(\tanh((\eta_i - R)/ l_s) - 1 \bigr) \, ,
\label{eq:6b}
\end{eqnarray}

previously introduced in simulations of DNA dynamics \cite{collins,druk}, enhancing by $D_i f_s$ the height of the barrier below which the base pair is closed. $l_s$ defines the width of the hump that modifies the plateau of $V_M$. For $\eta_i - R \gg  l_s$,  the Morse plateau is recovered, the two strands get apart from each other and the hydrogen bond with the solvent is established. The factor $f_s$ accounts for the solvent effects on the pair dissociation energy and can be empirically related to the salt concentration $[Na^+]$ in the solvent \cite{santa,owc,i11}.

\textit{c)} a nonlinear stacking potential

\begin{eqnarray}
& &V_S [\eta_i,\eta_{i-1}]=\, K \cdot G_{i, i-1} \cdot \bigl(\eta_i - \eta_{i-1}  \bigr)^2 \, \nonumber
\\
& &G_{i, i-1}=\,1 + \rho_{i, i-1} \exp\bigl[-\alpha_{i, i-1}(\eta_i + \eta_{i-1}  - 2R)\bigr]
\, , \nonumber
\\
\label{eq:6c}
\end{eqnarray}

first proposed in \cite{pey2}, measuring the coupling along the molecule stack between neighboring bases. Here, the harmonic force constant $K$ is assumed homogeneous while heterogeneity is introduced in the anharmonic parameters $\rho_{i, i-1}$ and $\alpha_{i, i-1}$ whose physical meaning is the following: if  $\eta_i - R < \alpha_{i, i-1}^{-1}$ for all \textit{ bps}, the double helix is intact and the effective coupling is $K \cdot (1 + \rho_{i, i-1})$. When a fluctuation occurs for the \textit{i-th bp} such that $\eta_i - R > \alpha_{i, i-1}^{-1}$ then the hydrogen bond loosens, the coupling drops to $K$ and the base moves out of the stack. This produces a fluctuational opening which may propagate cooperatively along the helix forming a bubble \cite{zocchi04}.
The conditions $\alpha_{i, i-1} < b_i$ should be fulfilled to ensure that the stacking potential range is larger than that of the Morse potential.
Note that kinks formation is possible in the circular model as \textit{bps} in adjacent fluctuational orbits may not be aligned along the stack \cite{i13}.

The Morse parameters are consistent with those obtained by fitting the melting curves of short DNA sequences \cite{campa}.
We take effective pair dissociation energies, $D_{AT}=\,30meV$ and $D_{GC}=\,45meV$, which are above $k_B \tau$ at room temperature and account for the inter-strand electrostatic repulsion due to the negatively charged backbone phosphates \cite{barbi12}.
The inverse lengths, $b_{AT}=\,2.4{\mathring{A} }^{-1}$ and $b_{GC}=\,2.7{\mathring{A} }^{-1}$, are set according to the suggestions of a study \cite{zdrav} of the model potential at the light of mechanical unzippering experiments \cite{heslot}.
The solvent potential parameters are $l_s=\,0.5 \mathring{A}$ and $f_s=\,0.1$ hence, we simulate a sizeable counterion concentration in the solvent, $[Na^+] \sim 0.4M$.
Varying these values has some effect on the helical repeat \cite{volo97} but it does not change the trend of our results.

Instead, large indeterminacy persists for the intra-strand parameters $K$, $\rho$'s and $\alpha$'s reflecting the fact that experimental data for the effective stacking force constant differ by two orders of magnitude. Generally, experiments probing the local length scale point to a high flexibility for the helix and report low
stacking couplings \cite{wiggins,fenn,weber09} whereas measurements of collective excitations such as longitudinal acoustic phonons report high stacking force constants \cite{eijck}.
As our method models the interactions at the local level, a weak
$K=\,20meV {\mathring{A} }^{-2}$  is set both for AT- and GC- \emph{bps} \cite{weber13} and the effects of the nonlinear stacking on the helix conformations are checked by tuning
the $\rho$'s and $\alpha$'s  parameters.

Both have been assumed homogeneous in previous studies \cite{pey2,singh} with $\rho$ values ranging from $0.5$ to $50$ albeit for models which represent DNA as a ladder hence not accounting for the helicoidal geometry. There is however a strong interplay between stacking and twist \cite{i12} posing some constraints on the anharmonic parameters as it will be shown below.
Here we introduce three types of parameters thus modeling the essential heterogeneous stacking couplings which may exist along the molecule axis:
\textit{1)} $\rho_1$, $\alpha_1$, denote the intra-strand coupling whenever
two neighboring bases are any of: $A-A$, $T-T$, $A-T$;
\textit{2)} $\rho_2$, $\alpha_2$,  indicate any two neighboring bases along the stack of the type: $A-G$, $A-C$, $T-G$, $T-C$;
\textit{3)} $\rho_3$, $\alpha_3$,
indicate any two neighboring bases along the stack of the type: $G-G$, $C-C$, $G-C$. The strands polarity is not accounted for in this model.
The inequalities, $\rho_1 > \rho_2 > \rho_3$  and $\alpha_1 < \alpha_2 < \alpha_3$, are fulfilled in the simulations to attribute larger anharmonic effects to the $A-T$ stacking.

If an $A$ or $T$ base moves out of the stack due to a thermal fluctuation,  the entropic gain will be larger in the case that the neighboring base along the stack is also $A$ or $T$. In this case, the stacking coupling will drop from $K(1 + \rho_1)$ to $K$.

\section*{IV. Computational method}

Our path integral technique considers the \textit{bps} radial displacements $|\textbf{r}_i|$ as time dependent paths which can be expanded in Fourier series. One set of Fourier coefficients selects a point, which corresponds to a DNA molecule state, in the path configuration space. The computational problem amounts to building (and integrating over) an ensemble of distinct configurations for the system.  Such ensemble is selected consistently with the physical constraints stemming from the model potential. For instance, the hard core barrier of the Morse potential (simulating the repulsion between complementary strands) excludes those paths whose fluctuations, $\eta_i \ll R$, would produce very large terms in Eq.~(\ref{eq:6a}) which, in turn, would yield vanishing contributions to the partition function.
As the paths ensemble is rebuilt at any temperature, the \textit{bps} thermal fluctuations around the ground state are fully incorporated in the path integral method \cite{i11}.
The computation includes $\sim 10^6$ distinct paths for each base pair.
This size permits to achieve numerical convergence in the free energy   per particle (units $meV$) up to three decimal places. This suffices to obtain monotonic $h$ versus $\tau$  as it is expected on physical grounds.

The thermodynamics of the DNA circles is derived from the classical partition function which reads

\begin{eqnarray}
& &Z_C =\,\oint \mathfrak{D}r \sum_{\theta_S} \exp\Bigl\{- \beta A_C[r] \Bigr\}\,\, \nonumber
\\
& & A_C[r] \equiv \,  \sum_{i=\,1}^{N}  \Bigl[\frac{\mu}{2} \dot{\eta}_i^2  + V_M[\eta_i] + \,V_{sol}[\eta_i] +  V_S[\eta_i,\eta_{i-1} ] \Bigr] \, . \nonumber
\\
\label{eq:7}
\end{eqnarray}

$\beta=\,(k_B \tau )^{-1}$ and $k_B$ is the Boltzmann constant. $\mu =\, 300$ a.m.u. is the \textit{bp} reduced mass.  The measure $\mathfrak{D}r$ is a multiple integral over the path Fourier coefficients. The integrals require temperature dependent cutoffs that truncate the path configuration space excluding those path amplitudes which, as mentioned above, are not consistent with the model potential \cite{i11a}.
The sum over a set of  $\theta_S$ values weighs an ensemble of distinct rotational conformations as the twist of the first \textit{bp} in the sequence
affects the fluctuational amplitudes at the successive sites.

The base pair fluctuation is measured with respect to the ground state. If the fluctuation is larger than a threshold $\zeta$, the base pair is taken as open.
Then, the status of the \textit{i-th} base pair is defined, for a specific fluctuation, by the Heaviside function $H_i$

\begin{eqnarray}
H_i \equiv H_i \bigl( |\eta_i - R| - \zeta \bigr) \,
\label{eq:9a}
\end{eqnarray}

The choice of $\zeta$ carries unavoidably some arbitrariness although the probability profiles shown below are not qualitatively modified by tuning $\zeta$ in the range $[1-2]\mathring{A}$ \cite{zhang97,ares,erp2,ares1}.
$\zeta =\,1\mathring{A} $ is taken hereafter.
By Eq.~(\ref{eq:9a}), we build the bubbles $H_i^{[d]}$ made of $d$ consecutive open \textit{bps} along the molecule backbone. The $i-th$ base pair is the center of the bubble (if $d$ is odd)  or the closest to the center from the left in Eq.~(\ref{eq:10}) (if $d$ is even).
Thus $d$ denotes the size of the bubble whose probability is generally expected to be proportional to the number of $AT$\textit{ bps} inside the bubble itself \cite{rapti,metz10}.
This effect however may not be evident for the circles in Eq.~(\ref{eq:10}) due to the almost regular alternation of AT and GC \textit{bps} along the sequences.

The bubble probabilities in each minicircle, at any site and for any possible $d$, are derived by carrying out an ensemble average in the paths configuration space formally expressed as

\begin{eqnarray}
< H_i^{[d]} >=\, Z_C^{-1} \oint \mathfrak{D}r \sum_{\theta_S} H_i^{[d]} \exp\Bigl\{- \beta A_C[r]\Bigr\} \, . \,
\label{eq:9b}
\end{eqnarray}

For the formation of open bubbles is a $\tau $ dependent process, Eq.~(\ref{eq:9b}) should be computed at any $\tau $ \textit{after} determining the twisting conformation peculiar of
that specific temperature.

\section*{V.  Results and Discussion}

\subsection*{A. Helix Unwinding }

The free energy, $F=\,\beta^{-1}\ln Z_C$, is calculated in the temperature range $[300, 360]K$, with $1K$ step, for the minicircles in Eq.~(\ref{eq:10}).
A broad ensemble of twisting conformations is assumed tuning $Tw$ with $0.0125$ partition step.  By minimizing $F$ as a function of $Tw$,  the helical repeat of the equilibrium conformation is evaluated. The procedure is repeated at any $\tau $ in the range, thus monitoring the helix unwinding due both to the bending stress and to the thermal fluctuations.  The execution time for a simulation, e.g. for the second sequence in Eq.~(\ref{eq:10}), is about 15 hours on a workstation (Intel Xeon E5-1620 v2, 3.7GHz processor).

\begin{figure}
\includegraphics[height=7.5cm,width=9.5cm,angle=0]{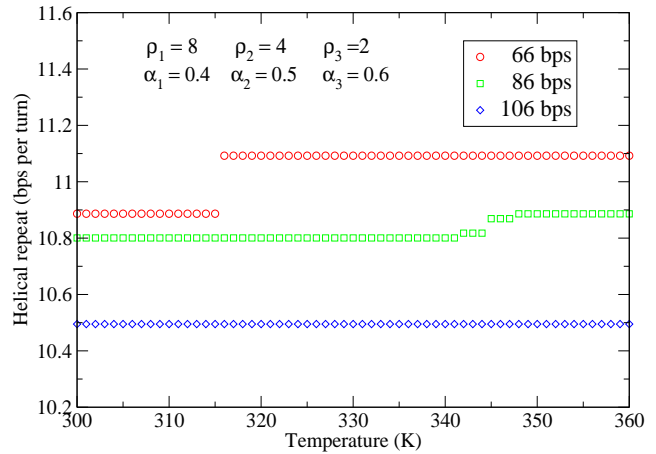}
\caption{\label{fig:1}(Color online) Helical repeats versus temperature for the three sequences in Eq.~(\ref{eq:10}). $\rho$'s parameters are dimensionless. $\alpha$'s parameters are in units {\AA}$^{-1}$.  }
\end{figure}

\begin{figure}
\includegraphics[height=7.5cm,width=9.5cm,angle=0]{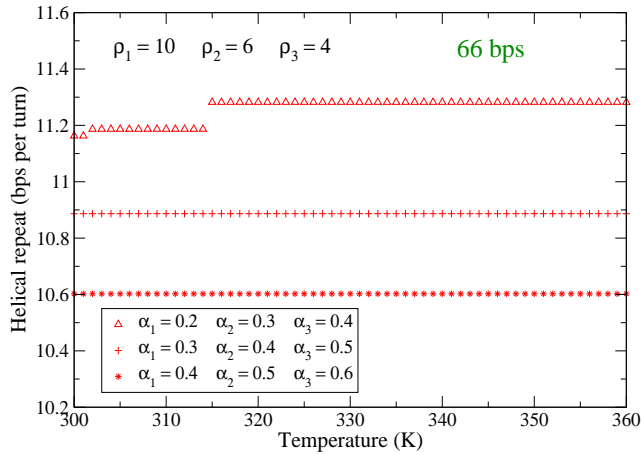}
\caption{\label{fig:2}(Color online) Helical repeats versus temperature for the third sequence in Eq.~(\ref{eq:10}) with three sets of $\alpha$'s parameters (in units {\AA}$^{-1}$).}
\end{figure}

The results are shown in Fig.~\ref{fig:1}. The potential parameters are common to all three minicircles to emphasize the effects of size and bending. $R$ is set for each minicircle so as to keep constant the rise distance between adjacent \textit{bps}, i.e. $\sim 3.4 \,\mathring{A} $.
While the largest circle shows no unwinding, the helical repeat of the
\textit{86-bps} sequence increases quite smoothly at $\tau \simeq \,345K$. The shortest circle shows instead a sizeable and abrupt unwinding at $\tau =\,315K$ consistent with the experimental findings of ref.\cite{volo08}. Here we see the interplay between twisting and bending as a function of the minicircle size. In general, the helix untwists to release the bending stress associated to the closure of the sequence into a loop. The untwisting is however driven by thermal fluctuations which become more effective in shorter sequences. For the latter, the energetic cost to be bent into a loop is in fact higher. Accordingly the \textit{66-bps} circle, with bending angle of $\sim  6$ degrees, shows an unwinding pattern remarkably different from those of the largest circles.

These \textit{close to the experiment} results have been obtained by selecting a set of anharmonic parameters.
Varying such set, i.e. enhancing the $\rho$'s, even the \textit{66-bps} circle shows a flat $h$ throughout the whole $\tau $ range, see Fig.~\ref{fig:2}. Only by reducing  significantly the $\alpha$'s, with respect to the values in Fig.~\ref{fig:1}, the thermal effect on $h$ is recovered but, in this case, $h$ turns out to be larger than $11$ even at room temperature.
While other, currently unavailable, experimental data may provide more stringent tests for the potential parameters, $\rho$'s smaller than $10$ seem appropriate to predict both the thermal unwinding and the formation of fluctuational openings in our smallest circle.

\subsection*{B. Bubble Statistics }

If $h$ measures an average property of the molecule, it matters to know which sites in the sequence may better sustain thermal fluctuations and which ones are more susceptible to disruptions which cause the helix untwisting discussed so far. The Hamiltonian approach permits to tackle this issue.

\begin{figure}
\includegraphics[height=7.5cm,width=9.5cm,angle=0]{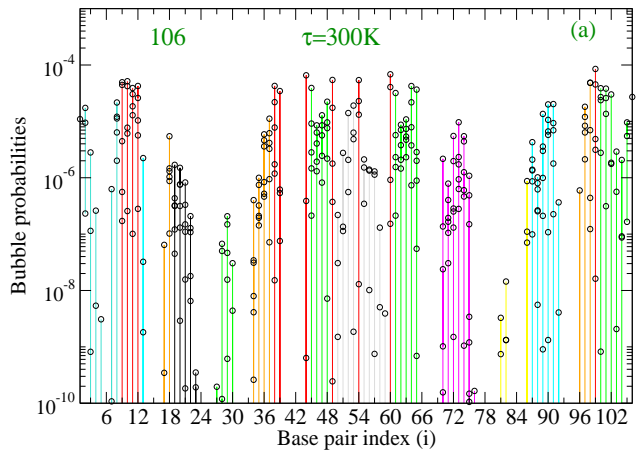}
\includegraphics[height=7.5cm,width=9.5cm,angle=0]{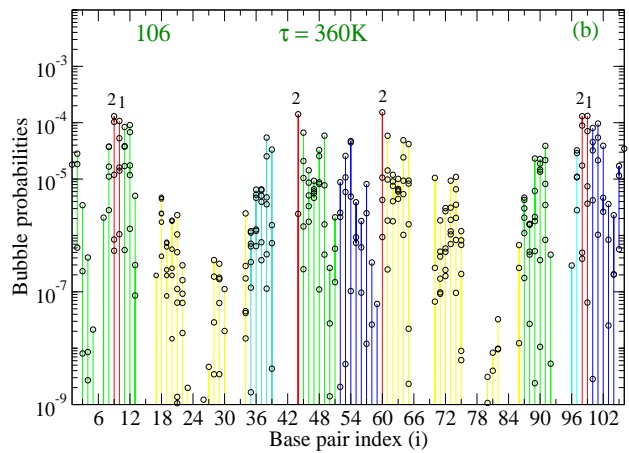}
\caption{\label{fig:3}(Color online) Bubble probabilities for the first sequence in Eq.~(\ref{eq:10}) computed by Eq.~(\ref{eq:9b}). (a) $\tau =\,300K$. (b) $\tau =\,360K$.
The circles mark the probabilities for bubbles of size $d$ whose centerpoint occurs at the base pair site $i$. At any site, probabilities larger than $10^{-9}$ are found only for bubbles with $d \leq 7$.
The red lines show, both in (a) and (b), the base pair sites at which bubbles are more likely to develop. However, only in (b) are bubble probabilities (slightly) larger than $10^{-4}$ at some sites. In these cases, the $d$ value found with higher probability is given on the top of the red line.}
\end{figure}

Fig.~\ref{fig:3} displays, for the longest minicircle in Eq.~(\ref{eq:10}), the probabilities for bubble formation computed, via Eq.~(\ref{eq:9b}), at the boundaries
of the $\tau$ range.
The parameters of Fig.~\ref{fig:1} are used hereafter.
In principle bubbles of any size compatible with the sequence length, if formed, can be detected by our algorithm. The $d$ sizes for bubbles occurring with probability at least $\sim  10^{ -4}$ are reported on top of the profiles. This order of magnitude is somewhat arbitrary reflecting the arbitrariness in the choice of $\zeta$, see Eq.~(\ref{eq:9a}).
As for short molecules, opening probabilities $\sim  10^{ -5}$ for AT \textit{bps} and $\sim  10^{ -6}$ for GC \textit{bps} have been estimated by a Ising-type statistical mechanics method \cite{krueg}, in fair agreement with measurements of exchange rates of DNA imino protons with solvent protons and NMR spectroscopy \cite{gueron,russu,russu1}. If analogous measurements had to become available for the minicircles, we may proceed to determine the specific $\zeta$ from the computation of the probability profiles.

No large probabilities are found at room temperature whereas, at $\tau =\,360K$ (see Fig.~\ref{fig:3}(b)), two single \textit{bp} fluctuations and four bubbles with $d=\,2$ show up with $< H_i^{[d]} > \, \sim 10^{ -4}$.  The $d$'s refer to the bubble sizes having largest probability but broader bubbles, centered on the same sites, are also found with lower probabilities.
Accordingly, the \textit{106-bps} circle displays a substantial stability at room temperature which is only marginally affected by some fluctuational openings appearing in the high $\tau $ regime.
This is consistent with the absence of helix unwinding pointed out in Fig.~\ref{fig:1}.

\begin{figure}
\includegraphics[height=7.5cm,width=9.5cm,angle=0]{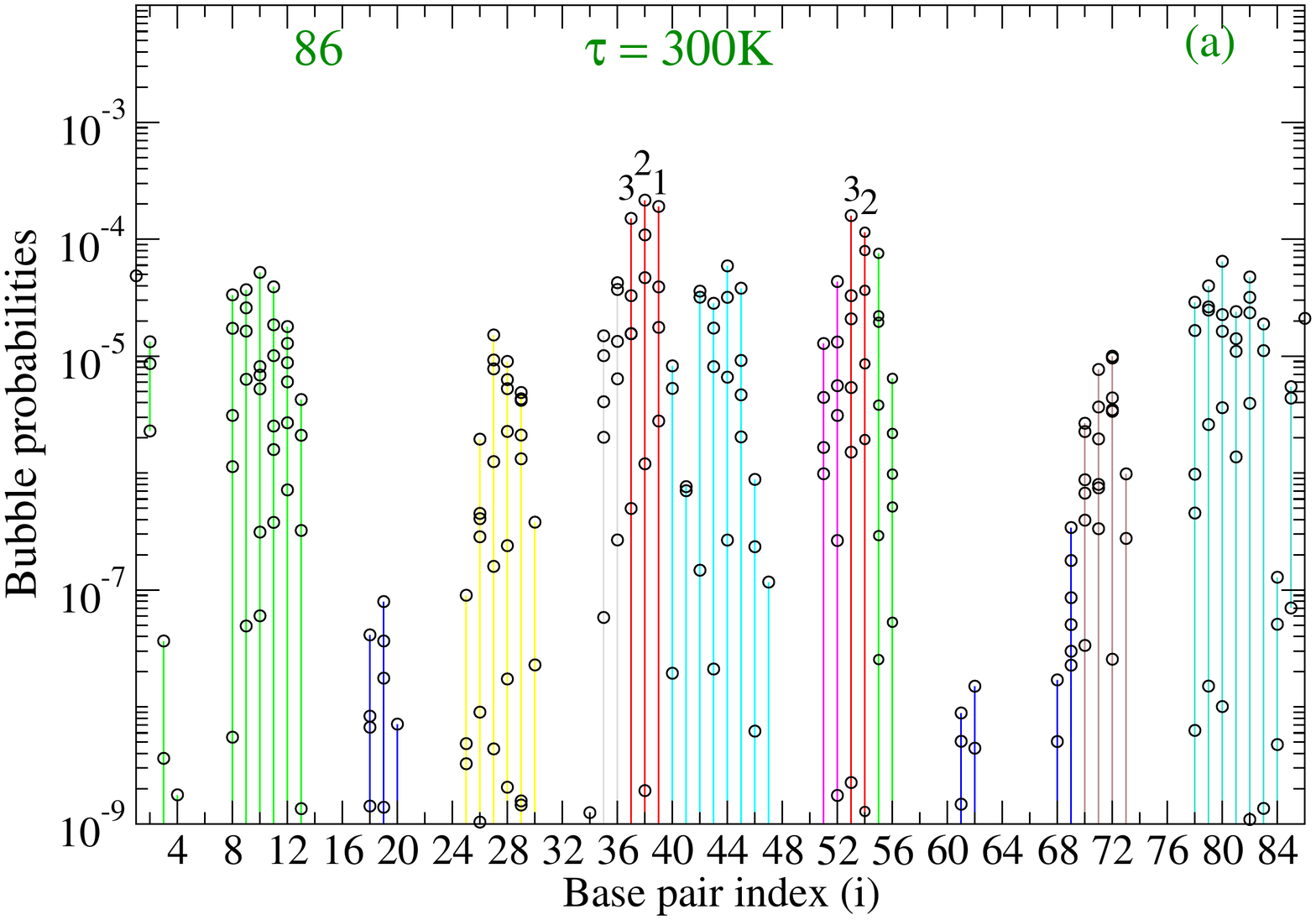}
\includegraphics[height=7.5cm,width=9.5cm,angle=-90]{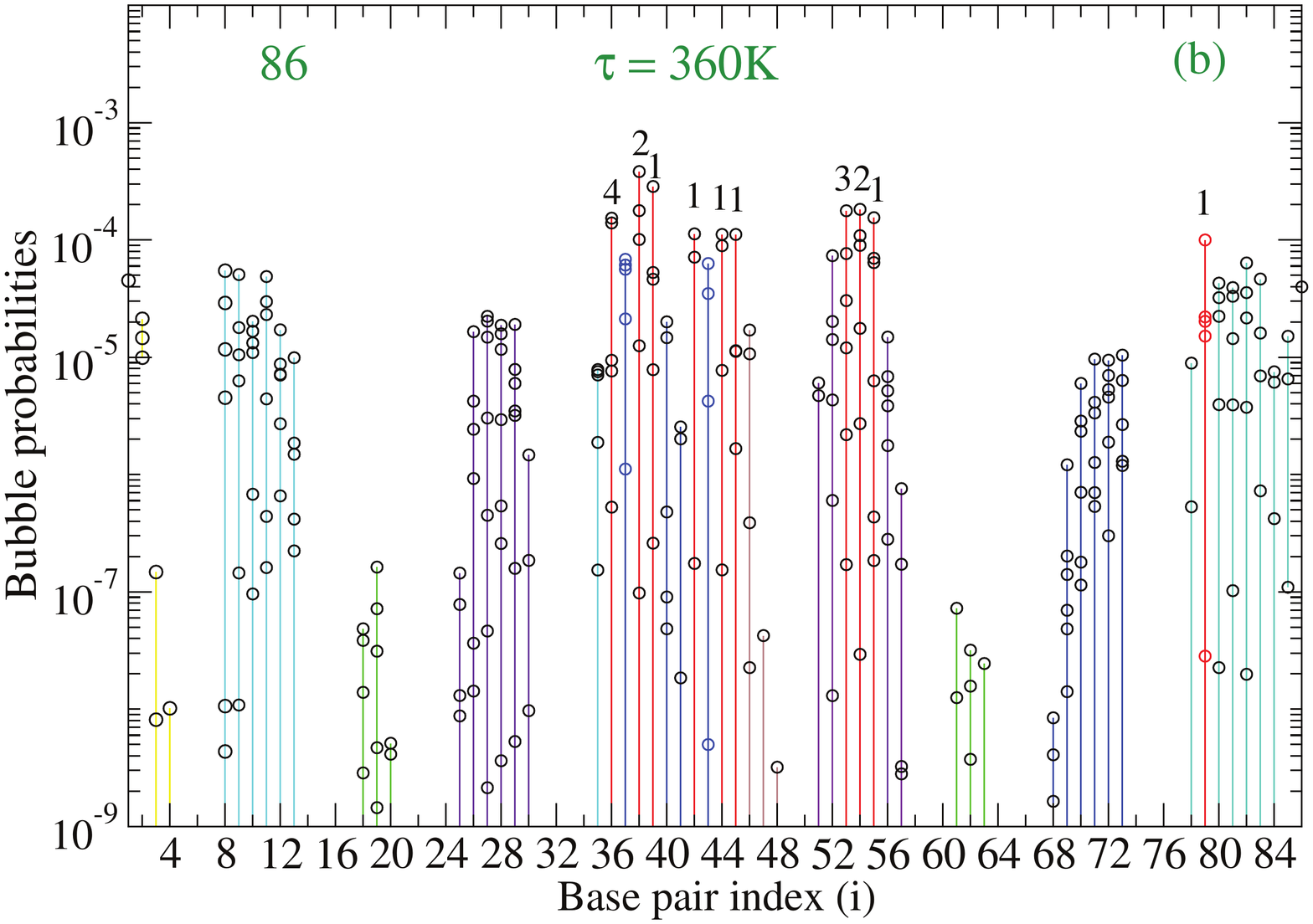}
\caption{\label{fig:4}(Color online) As in Fig.~\ref{fig:3} but for the second sequence in Eq.~(\ref{eq:10}). (a)  At base pair sites $i=\,37, 38, 39, 53, 54$, bubble probabilities are $ \geq 10^{-4}$. (b)  At sites $i=\,36, 38, 39, 42, 44, 45, 53, 54, 55, 79$, the probabilities are  $\geq 10^{-4}$.}
\end{figure}

The picture begins to change for the \textit{86-bps} minicircle whose bubble profiles are given in Fig.~\ref{fig:4}. Here two bubbles with $d=\,3$ and two bubbles with $d=\,2$ have probabilities larger than $10^{ -4}$ already at $\tau =\,300K$. Altogether $8$ \textit{bps} in the circle,  $9\%$ of the whole sequence, undergo large fluctuations (with relevant probability)  considering that the $i=\,38,\,39,\,54$ sites participate to two bubbles with neighboring centers. Some significant thermally driven \textit{bps} openings are detected at $\tau =\,360K$, see Fig.~\ref{fig:4}(b), which account for the $h$ increase found in Fig.~\ref{fig:1}. Note that the $d=\,4$ bubble is centered on a GC-\textit{bp} but it contains three AT-\textit{bps}.
At $i=26,\, 27$, probabilities are lower than $10^{-4}$ and therefore the size numbers have not been reported.
However, at these AT-sites, even large bubbles (up to $d=\,5$) may form with probabilities larger than $10^{-5}$.

\begin{figure}
\includegraphics[height=7.5cm,width=9.5cm,angle=0]{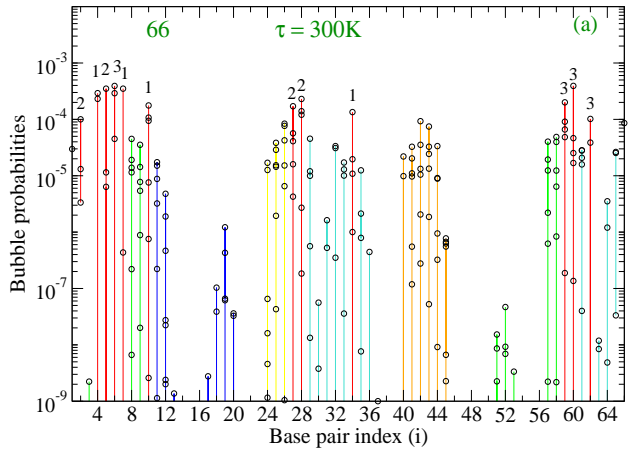}
\includegraphics[height=7.5cm,width=9.5cm,angle=-90]{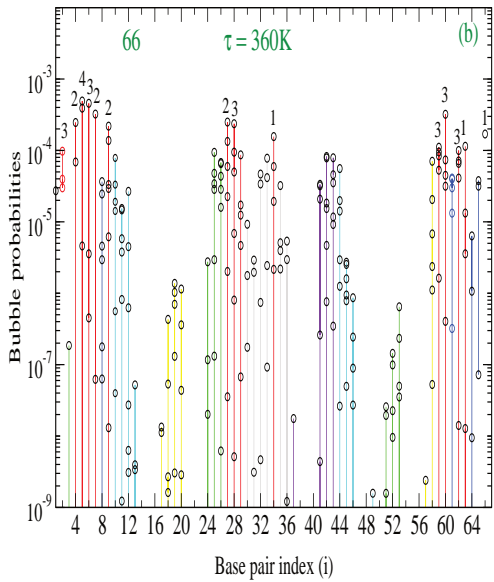}
\caption{\label{fig:5}(Color online)  As in Fig.~\ref{fig:3} but for the third sequence in Eq.~(\ref{eq:10}). (a)  At base pair sites $i=\,2, 4, 5, 6, 7, 10, 27, 28, 34, 59, 60, 62$, bubble probabilities are $\geq 10^{-4}$. (b)  At sites $i=\,2, 4, 5, 6, 7, 9, 27, 28, 34, 59, 60, 62, 63, 66$, the probabilities are $\geq 10^{-4}$.}
\end{figure}

A further spreading of bubbles is observed in the profiles for the \textit{ 66-bps} circle as displayed in Fig.~\ref{fig:5}. Now twelve bubbles appear at $\tau =\,300K$ involving $17$ \textit{bps} which make $ \sim 26\%$ of the whole sequence. Some of them, i.e. $i=\,5,\,6,\,7,\,27,\,28,\,59,\,60,\,61$ participate to two bubbles with centres at adjacent sites.  At the $i=\,60$ site, hosting a AT-bp, a bubble with $d=\,3$ occurs with the largest probability, $\sim 4 \cdot 10^{-4}$, but bubbles up to $d=\,6$ have also appreciable $< H_{ 60} ^{[d]} >$ values.
It is worth pointing out that bubble sizes of $ \sim 2 - 10$ \textit{bps} have in fact been detected by fluorescence correlation spectroscopy \cite{bonnet} in short DNA molecules at room $\tau $.
Moreover, there is a thermal effect on the bubbles profile, see Fig.~\ref{fig:5}(b),  with $20$ \textit{bps} having opening probabilities larger than $10^{-4}$ at the upper end of the $\tau $ range. For this minicircle however, the probabilities for disruption of the \textit{bps} bonds are quite high already at room temperature consistently with the large helix untwisting determined by minimization of the free energy.

Then, comparing the three minicircles bubble profiles at $\tau =\,300K$, it is found that \textit{bps} openings are much more likely to occur for the shortest sequence as a direct consequence of the wider bending angle,  $\phi_i \propto \, 1/N$, between fluctuational orbits of adjacent \textit{bps} along the molecule backbone. However, also the intermediate \textit{86-bps} minicircle with bending angle $\sim 4^o$ presents a few fluctuational openings at room temperature.
Starting from a Hamiltonian approach on the mesoscopic scale, our calculations are in line with the experiments of ref.\cite{volo08} and also confirm previous suggestions regarding the interplay of twisting and bending degrees of freedom based on analysis of the elastic free energy in DNA rings \cite{marko94}. We cannot determine a \textit{critical} circle size for the appearance of base pair disruptions as both their size and number do depend on the whole set of tunable model parameters, nonetheless our method can monitor the bubble development at specific sites and quantitatively predict the disruption probabilities as a function of the circle curvature.

\begin{figure}
\includegraphics[height=7.5cm,width=9.5cm,angle=-90]{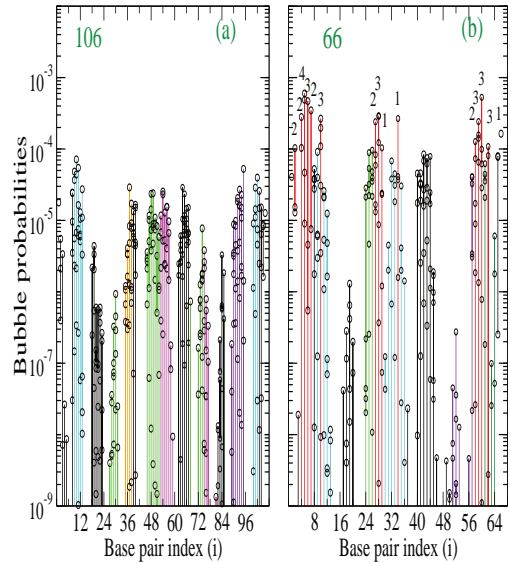}
\caption{\label{fig:6}(Color online)  Bubble probabilities for the first and third sequences in Eq.~(\ref{eq:10}), computed with the model parameters of Fig.~\ref{fig:1}, at $\tau =\,360K$.  The value $h=\,10.886$ is assumed for both sequences. This corresponds to the equilibrium helical repeat for the \textit{86-bps} sequence in Fig.~\ref{fig:4}(b).
(a) As the \textit{106-bps} sequence is undertwisted, the probabilities are lower than in Fig.~\ref{fig:3}(b). (b)  As the \textit{66-bps} is overtwisted,  the probabilities are enhanced with respect to Fig.~\ref{fig:5}(b).}
\end{figure}

The bubble statistics presented so far has been obtained by assuming, for each sequence, its own equilibrium helical repeat given in Fig.~\ref{fig:1}. We may also constrain all sequences to get the same $h$ and re-derive the bubble profiles: Fig.~\ref{fig:6} presents the results obtained at $\tau =\,360K$. The equilibrium value  $h=\,10.886$, peculiar of the \textit{86-bps} sequence, is assumed to hold also for the \textit{106-bps} and for the \textit{66-bps} sequences.
Hence, the $106-bps$ sequence is now \textit{undertwisted} whereas the $66-bps$ sequence is \textit{overtwisted}, with respect to their respective equilibrium values.
As we have reduced the torsional stress in the $106-bps$ sequence, Fig.~\ref{fig:6}(a) consistently shows lower bubble probabilities with respect to the corresponding equilibrium profile presented in Fig.~\ref{fig:3}(b). On the other hand, due to the enhanced torsional stress, the bubble probabilities for the $66-bps$ sequence (Fig.~\ref{fig:6}(b)) are  higher than in the equilibrium case of Fig.~\ref{fig:5}(b) with a few more sites participating to the formation of fluctuational openings.
These findings indicate that bubble profiles for a specific sequence may be largely affected by the torsional stress applied on the helix.

Eventually we should mention that bubble statistics (for linear sequences) had previously been computed  by molecular dynamics and direct integration methods in the framework of mesoscopic Hamiltonian models but with contrasting results, see e.g. \cite{choi,erp2}. While these models miss to describe the sequence heterogeneity in the stacking potential, even more importantly, they do not account for the helical torsional stress which is instead crucially related to the process of bubble formation \cite{benham06,erp06}. 

Instead, the importance of the twisting has been recognized by a recent Hamiltonian study of closure times of denaturation bubbles in linear sequences \cite{palmeri13}. However the latter work, applying the Langevin equation to a DNA model made of two freely rotating chains, assumes the torsional modulus as a tunable \textit{input} parameter which vanishes in the bubble.
While the focus of our present research differs from that of Ref.\cite{palmeri13},  we also point out that the method here proposed offers a significant advancement in the modeling of the DNA properties. In fact, helix unwinding and bubble statistics have been consistently obtained by \textit{first principles}, that is applying the principle of minimal action and selecting at any temperature those path ensembles which minimize the action $A_C[r]$.  Accordingly, the helical repeat represents an \textit{output} of the calculation and it is found as an equilibrium property of the molecule after summing over all base pair thermal fluctuations. Furthermore, the bubble profiles are related, at any temperature, to the degree of torsional (and bending) stress of the sequence.

Although a specific issue referring to circular DNA's has been here addressed, the path integral method can be straightforwardly extended to deal with open ends linear sequences.

\section*{VI. Conclusions }

The formation of base pair fluctuational openings, related to the bending of the molecule axis, has been investigated for three DNA minicircles having different sizes but similar AT and GC contents. The effective mesoscopic Hamiltonian includes the hydrogen bonds for AT and GC pairs, a solvent potential and the intra-strand nonlinear stacking coupling between neighboring bases. Twisting and bending degrees of freedom are incorporated in the circular model which then accounts for the occurrence of kinks and base pair disruptions. Hydrogen bonds and stacking interactions have been modeled by a single set of parameters consistent with a large body of experiments although the parametrization of the nonlinear stacking potential deserves more accurate analysis.
The path integral computational method considers the inter-strand base pair fluctuations as time dependent paths with one point of the path configuration space representing a molecule conformation. After summing over a large ensemble of paths and minimizing the free energy, we derive the helix twisting conformation which more effectively releases the bending stress due to the closure of the sequence into a ring.
While the helix unwinding as a function of temperature is driven by base pair thermal fluctuations, the size of the circle, i.e. the bending angle between neighboring base pair orbits, crucially determines the bubble probability profiles at room temperature.
Being all model parameters but the size common to the three sequences, we find that base pair disruptions are much more likely to occur in the shortest minicircle, i.e. the \textit{66-bps} circle with bending angle of $\sim 6$ degrees, in agreement with the data of single-strand-specific endonucleases experiments.
Consistently, the room temperature helical repeat is significantly larger in the smallest minicircle.

Our algorithm can detect, on any site in the sequence, the formation of bubbles and determine their broadening as a function of temperature. In general, for circular sequences whose bubble probability profiles are experimentally known, the algorithm can predict the threshold for base pair fluctuations above which the base pairs break.
Fluctuational effects around the ground state, which are
all the more strong in short sequences, are thus incorporated in the path integral method to an extent which cannot be reached by conventional two state models for the base pairs.

The computed probability profiles show that the
\textit{106-bps} circle is essentially stable whereas, in the \textit{66-bps} circle, $26\%$ of the sequence undergoes sizeable fluctuational openings already at room temperature. An intermediate behavior is obtained for the \textit{86-bps} circle with precisely $9\%$ of the sequence taking part in bubbles with relevant probabilities.
Unlike the \textit{106-bps} and \textit{86-bps} circles, the \textit{66-bps} circle also displays a large untwisting at $\tau =\,315K$ suggesting that a strong bending of the molecule axis renders the helix more susceptible to thermal effects.

While the presented results point altogether to a gradual increase in the number of base pair disruptions by reducing the size of minicircles below \textit{100-bps}, there is good prospect that combined experimental and computational analysis of more sequences, mostly in the range $\sim $ \emph{[ 60 - 80]- bps}, may settle the question of the existence of a critical radius of DNA curvature for the appearance of \textit{bps} disruptions. These questions should be also investigated by varying the relative contents of AT and GC pairs keeping the circle size fixed.
The theoretical perspective can be further improved by assuming that, in principle, adjacent base pair fluctuational orbits along the stack may have variable bending angles
and next, by determining the equilibrium bending conformation for the helix.


\begin{thebibliography}{widest-label}

\bibitem{benight}
P.V. Riccelli, F. Merante, K.T. Leung, S. Bortolin, R.L. Zastawny,
R. Janeczko, A.S. Benight,  \textit{Nucleic Acids Res.}, 2001, \textbf{29}, 996-1004.

\bibitem{genzer}
A. Jayaraman, C.K. Hall, J. Genzer,  \textit{J. Chem. Phys.}, 2007, {\bf 127}, 144912.

\bibitem{santa}
J. SantaLucia,    \emph{Proc. Natl. Acad. Sci. USA}, 1998, \textbf{95}, 1460-1465.

\bibitem{blake}
R.D. Blake, S.G. Delcourt,   \emph{Nucleic Acids Res.}, 1998, {\bf 26}, 3323-3332.

\bibitem{owc}
R. Owczarzy, Y. You, B.G. Moreira, J.A. Manthey, L. Huang, M.A. Behlke, J.A. Walder,  \emph{Biochemistry}, 2004, \textbf{43}, 3537-3554.

\bibitem{rapti10}
M.R. Kantorovitz, Z. Rapti, V. Gelev, A. Usheva, \textit{Bioinformatics}, 2010, \textbf{11}, 604.


\bibitem{seeman}
N.C. Seeman,  \emph{J. Theor. Biol.}, 1982, \textbf{99}, 237-247.

\bibitem{rose}
K. Komiya, M. Yamamura, J.A. Rose,  \emph{Nucleic Acids Res.}, 2010, {\bf 38}, 4539-4546.

\bibitem{zhao}
Y.-X. Zhao, A. Shaw, X. Zeng, E. Benson, A.M. Nystr\"{o}m, B. H\"{o}gberg,  \emph{ACS Nano}, 2012, \textbf{6}, 8684-8691.

\bibitem{cha}
P.F. Xu, H. Noh, J.H. Lee, D.W. Domaille, M.A. Nakatsuka, A.P. Goodwin, and J.N. Cha,  \emph{Materials Today}, 2013, \textbf{16}, 290-296.

\bibitem{albu}
E.L. Albuquerque, U.L. Fulco, V.N. Freire, E.W.S. Caetano, M.L. Lyra, F.A.B.F. Moura, \textit{Phys. Rep.} in press.

\bibitem{crothers}
J.C. Marini, S.D. Levene, D.M. Crothers, P.T. Englund,  \emph{Proc. Natl. Acad. Sci. USA}, 1982, \textbf{79}, 7664-7668.

\bibitem{shore}
D. Shore, J. Langwoski, R.L. Baldwin,  \emph{Proc. Natl. Acad. Sci. USA}, 1981, \textbf{78}, 4833-4837.

\bibitem{ohyama}
T. Ohyama,  \emph{BioEssays}, 2001, \textbf{23}, 708-715.

\bibitem{choi}
C.H. Choi, G. Kalosakas, K.{\O}. Rasmussen, M. Hiromura, A.R. Bishop and A. Usheva, \textit{Nucleic Acids Res.}, 2004, \textbf{32}, 1584-1590.

\bibitem{cloutier}
T.E. Cloutier, J. Widom,  \emph{Mol. Cell}, 2004, \textbf{ 14}, 355-362.

\bibitem{stock}
H. Jacobson, W.H. Stockmayer,   \emph{J. Chem. Phys.}, 1950, \textbf{18}, 1600-1606.

\bibitem{shimada}
J. Shimada, H. Yamakawa,  \emph{Macromolecules}, 1984, \textbf{17}, 689-698.

\bibitem{yan04}
J. Yan, J.F. Marko,  \textit{Phys. Rev. Lett.}, 2004, \textbf{ 93},  108108.


\bibitem{crick}
F.H. Crick, A. Klug,  \emph{Nature}, 1975, \textbf{255}, 530-533.

\bibitem{maddocks}
F. Lank\v{a}s, R. Lavery, J.H. Maddocks,  \emph{Structure}, 2006, \textbf{14}, 1527-1534.

\bibitem{volo05}
Q. Du, C. Smith, N. Shiffeldrim, M. Vologodskaia, A. Vologodskii,  \emph{Proc. Natl. Acad. Sci. USA}, 2005, \textbf{102}, 5397-5402.


\bibitem{zocchi13}
D.S. Sanchez, H. Qu, D. Bulla, G. Zocchi, \emph{Phys. Rev. E}, 2013, {\bf 87}, 022710.

\bibitem{volo08}
Q. Du,  A. Kotlyar, A. Vologodskii,   \emph{Nucl. Acids Res.}, 2008, \textbf{ 36},  1120-1128.

\bibitem{volo09}
X. Zheng, A. Vologodskii,  \emph{Biophys. J.}, 2009, \textbf{96}, 1341-1349.

\bibitem{vafa}
R. Vafabakhsh, T. Ha,   \textit{Science}, 2012, \textbf{ 337}, 1097-1101.

\bibitem{volo13}
A. Vologodskii, M.D. Frank-Kamenetskii,  \emph{Nucl. Acids Res.}, 2013, \textbf{ 41},  6785-6792.

\bibitem{i11}
M. Zoli, \emph{J. Chem. Phys.}, 2011, \textbf{135}, 115101.

\bibitem{duguet}
M. Duguet,  \emph{Nucleic Acids Res.}, 1993, \textbf{21}, 463-468.

\bibitem{depew}
R.E. Depew, J.C. Wang,
\textit{Proc. Natl. Acad. Sci. U.S.A.}, 1975, \textbf{72}, 4275-4279.

\bibitem{benham93b}
W.R. Bauer, C.J. Benham,    \emph{J. Mol. Biol.}, 1993,  \textbf{234}, 1184-1196.

\bibitem{benham99}
R.M. Fye,  C.J. Benham,  \textit{Phys. Rev. E}, 1999, {\bf 59}, 3408-3426.

\bibitem{metz12}
J. Adamcik,  J.-H. Jeon,   K.J. Karczewski,  R. Metzler,  G. Dietler,
\textit{Soft Matter}, 2012, \textbf{8}, 8651-8658.

\bibitem{bates}
A.D. Bates, A. Maxwell,  \emph{DNA Topology}, Oxford University Press, Oxford, UK, 2005.

\bibitem{marko95}
J.F. Marko, E.D. Siggia,  \textit{Phys. Rev. E},  1995, {\bf 52}, 2912-2938.

\bibitem{shore83}
D. Shore, R.L. Baldwin,  \textit{J. Mol. Biol.},  1983, \textbf{170}, 983-1007.

\bibitem{pey2}
T. Dauxois, M. Peyrard, A.R. Bishop,    \emph{Phys. Rev. E}, 1993, \textbf{47},  R44-47.

\bibitem{collins}
F. Zhang, M.A. Collins,   \emph{Phys. Rev. E}, 1995, {\bf 52}, 4217-4224.

\bibitem{druk}
K. Drukker, G. Wu, G.C. Schatz,  \emph{J. Chem. Phys.}, 2001, \textbf{114}, 579-590.

\bibitem{zocchi04}
Y. Zeng,  A. Montrichok,  G. Zocchi,   \emph{J. Mol. Biol.},  2004, \textbf{339}, 67-75.

\bibitem{i13}
M. Zoli, \emph{J. Chem. Phys.},  2013, {\bf 138}, 205103.

\bibitem{campa}
A. Campa, A. Giansanti, \textit{Phys. Rev. E},  1998, \textbf{58}, 3585-3588.


\bibitem{barbi12}
P. Carrivain, A. Cournac, C. Lavelle, A. Lesne, J. Mozziconacci, F. Paillusson, L. Signon, J.M. Victor, M. Barbi,
\textit{Soft Matter}, 2012, \textbf{8}, 9285-9301.

\bibitem{zdrav}
S. Zdravkovi\'{c}, M.V. Satari\'{c},
\emph{Phys. Rev. E}, 2006, {\bf 73},  021905.

\bibitem{heslot}
U. Bockelmann, B. Essevaz-Roulet, F. Heslot,  \emph{Phys. Rev. Lett.}, 1997, \textbf{79}, 4489-4492.

\bibitem{volo97}
V.V. Rybenkov, A.V. Vologodskii, N.R. Cozzarelli,
\textit{Nucl. Acids Res. },  1997, \textbf{ 25}, 1412-1418.

\bibitem{wiggins}
P.A. Wiggins, T. van der Heijden, F. Moreno-Herrero, A. Spakowitz, R. Phillips, J. Widom, C. Dekker,  P.C. Nelson,
\textit{Nature Nanotech.},  2006, \textbf{ 1}, 137-141.

\bibitem{fenn}
R. S. Mathew-Fenn, R. Das, and P. A. B. Harbury,  \textit{Science},  2008, \textbf{322}, 446-449.

\bibitem{weber09}
G. Weber, J.W. Essex, C. Neylon,   \textit{Nat. Phys.}, 2009, \textbf{5}, 769–773.

\bibitem{eijck}
L. van Eijck, F. Merzel, S. Rols, J. Ollivier, V.T. Forsyth, M.R. Johnson,  \textit{Phys. Rev. Lett.}, 2011, \textbf{ 107},  088102.

\bibitem{weber13}
G. Weber, \textit{Nucl. Acids Res.}, 2013, \textbf{41}, e30.


\bibitem{singh}
S. Srivastava, N. Singh,   \emph{J. Chem. Phys.}, 2011, \textbf{134},  115102.

\bibitem{i12}
M. Zoli,  \emph{J. Phys.: Condens. Matter}, 2012, {\bf 24},  195103.

\bibitem{i11a}
M. Zoli, \emph{Eur. Phys. J. E},  2011, {\bf 34}, 68.

\bibitem{zhang97}
Y. Zhang, W.M. Zheng, J.X. Liu, Y.Z. Chen,
\textit{ Phys.\ Rev.\ E},  1997, {\bf 56}, 7100-7115.

\bibitem{ares}
S. Ares, N.K. Voulgarakis, K.{\O}. Rasmussen, A.R. Bishop, \textit{ Phys.\ Rev.\ Lett.},  2005, {\bf 94}, 035504.

\bibitem{erp2}
T.S. van Erp, S. Cuesta-Lopez, M. Peyrard,  \textit{Eur. Phys. J. E},  2006, {\bf 20}, 421-434.

\bibitem{ares1}
G. Kalosakas, S. Ares,  \textit{J. Chem. Phys.},  2009, {\bf 130}, 235104.

\bibitem{rapti}
Z. Rapti, A. Smerzi, K.{\O}. Rasmussen, A.R. Bishop, C. H. Choi, A. Usheva,  \textit{ Phys.\ Rev.\ E},  2006, {\bf 73}, 051902.

\bibitem{metz10}
J.H. Jeon, J. Adamcik, G. Dietler, R. Metzler,
\textit{Phys.\ Rev.\ Lett. }, 2010, {\bf 105}, 208101.

\bibitem{krueg}
A. Krueger, E. Protozanova, M.D. Frank-Kamenetskii, \textit{Biophys. J.}, 2006, {\bf 90}, 3091-3099.

\bibitem{gueron}
M. Gu\'{e}ron, M. Kochoyan, J.L. Leroy,  \textit{Nature},  1987, \textbf{ 328}, 89-92.

\bibitem{russu}
C. Chen, I.M. Russu,  \emph{Biophys. J.}, 2004, \textbf{87}, 2545-2551.

\bibitem{russu1}
D. Coman, I.M. Russu,  \emph{Biophys. J.},  2005, \textbf{89}, 3285-3292.

\bibitem{bonnet}
G. Altan-Bonnet,  A. Libchaber, O. Krichevsky,   \emph{Phys. Rev. Lett.}, 2003, \textbf{90}, 138101.

\bibitem{marko94}
J.F. Marko, E.D. Siggia,   \emph{Macromolecules}, 1994, {\bf 27},  981-988.

\bibitem{benham06}
C.J. Benham, R.R.P. Singh, \textit{Phys. Rev. Lett.},  2006, {\bf 97}, 059801.

\bibitem{erp06}
T.S. van Erp, S. Cuesta-Lopez, J.G. Hagmann, M. Peyrard,  \textit{Phys. Rev. Lett.},  2006, {\bf 97}, 059802.


\bibitem{palmeri13}
A.K. Dasanna, N. Destainville, J. Palmeri, M. Manghi, \textit{Phys. Rev. E }, 2013, \textbf{87}, 052703.


\end{thebibliography}
\end{document}